\documentclass[fleqn,usenatbib]{mnras}
\usepackage[T1]{fontenc}
\usepackage{ae,aecompl}
\usepackage{graphicx}   
\usepackage{amsmath}    
\usepackage{amssymb}    
\usepackage{float}

\newcommand{\ntab}[2]{ \multicolumn{1}{#1}{#2} }
\newcommand{\nntab}[2]{ \multicolumn{2}{#1}{#2} }

\usepackage{xcolor}
\definecolor{Dred}{rgb}{0.312,0.070,0.070}
\definecolor{Dblue}{rgb}{0.070,0.070,0.312}

\newcounter{note}
\let\oldmarginpar\marginpar
\renewcommand\marginpar[1]{\-\oldmarginpar[\raggedleft\footnotesize #1]%
{\raggedright\footnotesize #1}}
\newcommand{\Note}[1]{\Rdb{#1}{\addtocounter{note}{1}%
\marginpar{\small\underline{\Rdb{Comm \arabic{note}}}}}}
\newcommand{\note}[1]{\Rdb{#1}}
\newcommand{\beq}{ \begin{eqnarray} }
\newcommand{\eeq}[1]{\label{#1}\end{eqnarray}}
\newcommand{\eeqn}{ \nonumber \end{eqnarray} }
\newcommand{\Frac}[2]{\frac{\displaystyle\strut #1}{\displaystyle\strut #2} }
\newcommand{\vex}{\vspace{1ex}}

\renewcommand{\note}{}\renewcommand{\Note}{} 

\newcommand{\timezone}{-0400}
\newcommand{\Number}[1]{\ifnum#1<10\relax0\number#1\else\number#1\fi}
\newcommand{\isodate}{
\count151=\time
\divide\count151 by 60
\count151=\count151
\multiply\count151 by 60
\count152=\time
\advance\count152 by -\count151
\divide\count151 by 60
\count152=\count151
\multiply\count151 by 60
\count153=\time
\advance\count153 by -\count151
\Number{\year}.\Number{\month}.\Number{\day}--\Number{\count152}:\Number{\count153} \enskip \timezone
}
\definecolor{OliveGreen}{rgb}{0,0.6,0}

\title[Systematic differences VLBI/Gaia]
      {\vspace{-4ex} \LARGE\bf 
         A quantitative analysis of systematic differences in positions and 
         proper motions of Gaia DR2 with respect to VLBI
      }
\author[Petrov et al.]{
L.~Petrov$^{1}$\thanks{E-mail: Leonid.Petrov@nasa.gov},
Y.~Y.~Kovalev$^{2,3,4}$ and
A.~V.~Plavin$^{2,3}$
\\
$^1$NASA Goddard Space Flight Center, Greenbelt, ND 20771, USA \\
$^2$Astro Space Center of Lebedev Physical Institute, Profsoyuznaya 84/32, 
   117997 Moscow, Russia\\
$^3$Moscow Institute of Physics and Technology, Dolgoprudny, Institutsky per., 9, Moscow, Russia \\
$^4$Max-Planck-Institut f\"ur Radioastronomie, Auf dem H\"ugel 69, 
   53121 Bonn, Germany
}

\date{Accepted 12 October 2018. Received 27 September 2018; in original form 15 August, 2018}

\pubyear{2018}

\begin{document}
\volume{}
\pubyear{2018}
\setcounter{page}{1}
\label{firstpage}
\pagerange{\pageref{firstpage}--\pageref{lastpage}}

\maketitle

\begin{abstract}
   We have analyzed the differences in positions of 9081 matched sources
between the Gaia DR2 and VLBI catalogues. The median position uncertainty 
of matched sources in the VLBI catalogue is a factor of two larger than 
the median position uncertainty in the Gaia DR2. There are 9\% matched 
sources with statistically significant offsets between both catalogues.
We found that reported positional errors should be re-scaled by a factor 
of 1.3 for VLBI and 1.06 for Gaia, and in addition, Gaia errors should be 
multiplied by the square root of chi square per degree of freedom in order 
to best fit the normalized position differences to the Rayleigh distribution. 
We have established that the major contributor to statistically significant 
position offsets is the presence of optical jets. Among the sources for 
which the jet direction was determined, the position offsets are parallel 
to the jet directions for 62\% of the outliers. Among the matched sources 
with significant proper motion, the fraction of objects with proper motion 
directions parallel to jets is a factor of 3 greater than on average. Such 
sources have systematically higher chi square per degree of freedom. We 
explain these proper motions as a manifestation of the source position jitter 
caused by flares that we have predicted earlier. Therefore, the assumption 
that quasars are fixed points and therefore, differential proper motions 
determined with respect to quasar photocenters can be regarded as absolute 
proper motions, should be treated with a great caution.
\end{abstract}

\begin{keywords}
galaxies: active~--
galaxies: jets~--
quasars: general~--
radio continuum: galaxies~--
astrometry: reference systems
\end{keywords}

\section{Introduction}

  Since 1980s very long baseline interferometry (VLBI) has been the most 
accurate absolute astrometry technique. The accuracy of VLBI absolute 
positions can reach the 0.1~mas level. With few exceptions, VLBI is able 
to provide absolute positions of only active galactic nuclei (AGNs). 
In 2016, the Gaia Data Release~1 (DR1) \citep{r:gaia_dr1} ushered an 
emergence of the technique that rivals VLBI in accuracy. A quick analysis 
by \citet{r:gaia_icrf2} found that in general, the differences between 
common AGNs in VLBI and Gaia DR1 catalogues are close to their 
uncertainties, except for a 6\% of common objects. \citet{r:gaia_icrf2} 
claims that ``individual examination of a number of these cases shows 
that a likely explanation for the offset can often be found, for example 
in the form of a bright host galaxy or nearby star''. They conclude 
(page 13) that ``the overall agreement between the optical and radio 
positions is excellent''. We see it differently. If two independent 
observing campaigns produced small (negligible) differences, that also 
implies that the contribution of a new campaign is also small (negligible) 
with respect to what has been known before. Science does not emerge from 
agreements. It emerges from disagreements. Therefore, we focused our 
analysis on differences between VLBI and Gaia AGN positions.

  Our analysis of Gaia DR1 confirmed the existence of a population of sources 
with statistically significant VLBI/Gaia offsets \citep{r:gaia1}. We 
found that such factors as the failures in quality control in both VLBI and 
Gaia, blended nearby stars, or bright host galaxies can account at maximum 
for 1/3 of that population. This analysis, as well as recent works of others 
\citep{r:gaia_icrf2,r:mak17,r:Frouard18,r:Liu18a,r:Liu18b,r:Liu18c}, 
used arc lengths of VLBI/Gaia differences. Including the second dimension, 
the position angle of VLBI/Gaia offsets, resulted in a breakthrough. Though 
the distribution of the position angles counted from the declination axis 
turned out to be close to uniform, the distribution of the position angles 
with respect to the jet direction determined from analysis of VLBI images of 
matched sources revealed a strong anisotropy \citep{r:gaia2}: the offsets
have a preferable direction along the jet, and at a smaller extent in the
direction opposite to the jet. We interpret it as a manifestation of a 
presence of optical jets at scales finer than the Gaia point spread 
function (PSF), i.e., 100--300~mas. Known optical jets in AGNs resolved
with Hubble Space Telescope are cospatial \citep{r:gabuzda2006,r:perlman2010,
r:meyer18_m84}. Even in that case there will be position differences. 
It was emphasized in \citep{r:gaia3} that the response to an extended 
structure of a power detector used by Gaia and an interferometer that 
records voltage is fundamentally different. The Gaia positions
correspond to the location of the optical centroid, while the VLBI positions
are associated to the most compact and bright feature at the jet 
base. Therefore, the physical meaning of a VLBI/Gaia offset is 
a displacement of the optical centroid with respect to the jet base.

  In April 2018, the Gaia DR2 was published \citep{r:gaia_dr2}. It has 
48\% more sources than Gaia DR1 and a significantly higher accuracy.
\mbox{\citet{r:gaia_dr2_crf}} reported that in general, the VLBI/Gaia DR2 
differences are small with some exceptions. They set out five reasons
for discrepancies (page 10): 1)~real offsets between the centres of
emission at optical and radio wavelengths; 2)~error in matching 
VLBI and Gaia objects; 3)~an extended galaxy around the quasar; 
4)~double or lensed quasars; or 5)~simply statistical outliers.
The presence of optical jets was not put in the list as a likely 
explanation.

  In \citet{r:gaia3} we examined the consequences of our interpretation
of the  VLBI/Gaia offsets due to the presence of optical jets. Among others,
we made two predictions: 1)~``further improvement in the position 
accuracy of VLBI and Gaia will not result in a reconciliation of 
radio and optical positions, but will result in improvement of the 
accuracy of determination of these position differences'', 
2)~``we predict a jitter in the Gaia centroid position estimates for 
radio-loud AGNs''. Since the Gaia DR2 accuracy is noticeably 
better than the Gaia DR1 accuracy, this motivated us to extend 
our previous analysis to the Gaia DR2 and check whether these predictions 
came true. To answer the question what is the most significant contributor 
to systematic position differences is the goal of this article.

\section{Comparison of VLBI/Gaia positions}

\begin{figure}
    \centerline{\includegraphics[width=0.48\textwidth]{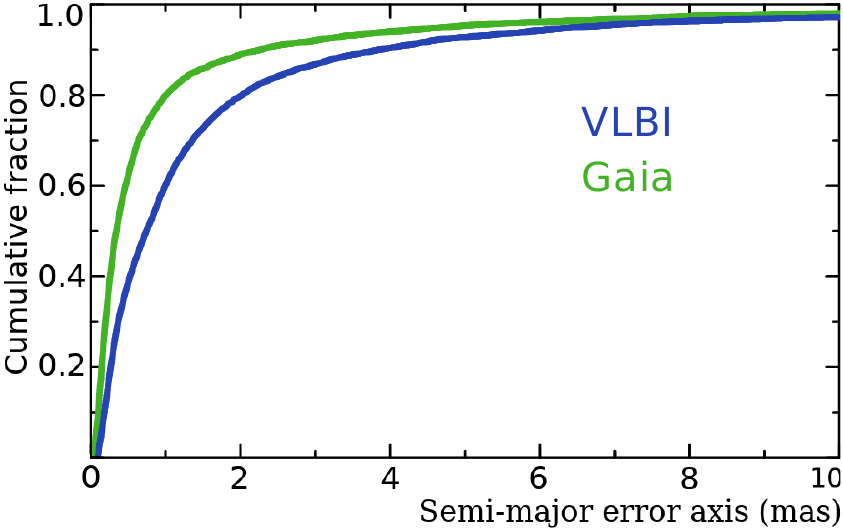}}
    \caption{The cumulative distribution function of semi-major axes of error
             ellipses $P(\sigma_{\rm maj}<a)$: green (upper) curve for Gaia and 
             blue (low) curve for VLBI.}
    \label{f:cumul}
\end{figure}

   We matched the Gaia DR2 catalogue of 1,692,919,135 objects against the 
Radio Fundamental Catalogue rfc\_2018b (L. Petrov and Y.Y.~Kovalev in 
preparation, 2018)\footnote{Available online at 
\href{http://astrogeo.org/rfc}{http://astrogeo.org/rfc}} (RFC) of 15,155
sources. The RFC catalogue is derived using all VLBI observations
under astrometric programs publicly available by August 01 2018.
We used the same procedure of matching Gaia objects against the VLBI 
catalogue described in detail in \citet{r:gaia1} and got 9081 matches 
with the probability of false association below the $2 \cdot 10^{-4}$ level. 
The immediate comparison of formal uncertainties {\it among matches} showed 
that the Gaia uncertainties are smaller (see Figure~\ref{f:cumul}). The median 
of VLBI semi-major axes of error ellipses is 0.74~mas against 0.34~mas
for Gaia. Although VLBI can reach accuracies of 0.1~mas in absolute 
positions of strong sources, the majority of the sources were observed only 
once for 60~seconds, which is insufficient to derive their positions with 
that level of accuracy. The Gaia uncertainties of matches are roughly twice
smaller than the VLBI uncertainties, though this statement cannot be 
generalized to the entire Gaia or VLBI catalogues.

  Among 9081 matches, radio images at milliarcsecond resolution for 8143 sources
are available. Using them, we have evaluated the jet directions 
for 4030 sources, i.e., for one half of the sample. We removed 48 sources 
that include 13 radio stars, 1 supernova remnant in the nearby 
star-forming galaxy, 10 gravitational lenses, and 24 double objects.

\subsection{Analysis of VLBI/Gaia position angles with respect to the 
jet direction}

  We examined the arc lengths $a$ between the VLBI and Gaia source position 
estimates as well as the position angles $\phi$ of Gaia positions with respect 
to VLBI positions counted counter-clockwise with respect to the declination 
axis. Using reported position uncertainties and correlations between right 
ascensions and declinations, we computed the semi-major and semi-minor axes 
of error ellipse, as well as their position angles $\theta$ for both VLBI 
and Gaia position estimates. Then, assuming VLBI and Gaia errors are 
independent, we computed the uncertainties of arc lengths $\sigma_a$  
and position angles $\sigma_\phi$ in the linear approximation:
 
\beq
  \hspace{-1em}
  \begin{array}{l@{\:}c@{\:}l}
     \sigma^2_a & = & \note{           
                      \Frac{1 + \tan^2(\theta_v - \phi)}
                           {1 + \frac{\sigma^2_{v,\rm maj}}{\sigma^2_{v,\rm min}} \tan^2(\theta_v - \phi)} \, \sigma^2_{v,\rm maj} \, 
                      \: + \:
                      \Frac{1 + \tan^2(\theta_g - \phi)}
                           {1 + \frac{\sigma^2_{g,\rm maj}}{\sigma^2_{g,\rm min}} \tan^2(\theta_g - \phi)} \, \sigma^2_{g,\rm maj} \, 
                      }
     \vex \\
     \sigma^2_\phi & = & \phantom{2} \Delta(\alpha_g - \alpha_v)^2 
                         (\sigma_{v,\delta}^2 + \sigma_{g,\delta}^2) \cos^2\delta_v/a^4 \: + \\ &&
                         \phantom{2} \Delta(\delta_g - \delta_v)^2 
                         (\sigma_{v,\alpha}^2 + \sigma_{g,\alpha}^2) \cos^2\delta_v/a^4 \: - \\ &&
                         2 \Delta(\alpha_g - \alpha_v) \Delta(\delta_g - \delta_v) \times    \\ &&
                         \phantom{2 \Delta}
                         ( {\rm Corr}_v \sigma_{v,\alpha} \sigma_{v,\delta} +
                           {\rm Corr}_g \sigma_{g,\alpha} \sigma_{g,\delta} ) \cos^2\delta_v/a^4 ,
  \end{array}
\eeq{e:e1}
\par\noindent
   where Corr is the correlation between right ascension and declination and 
the uncertainties in right ascensions are without the $\cos\delta$ factor. 
Labels $v$ and $g$ stand for VLBI and Gaia respectively.

\begin{figure}
    \centerline{\includegraphics[width=0.48\textwidth]{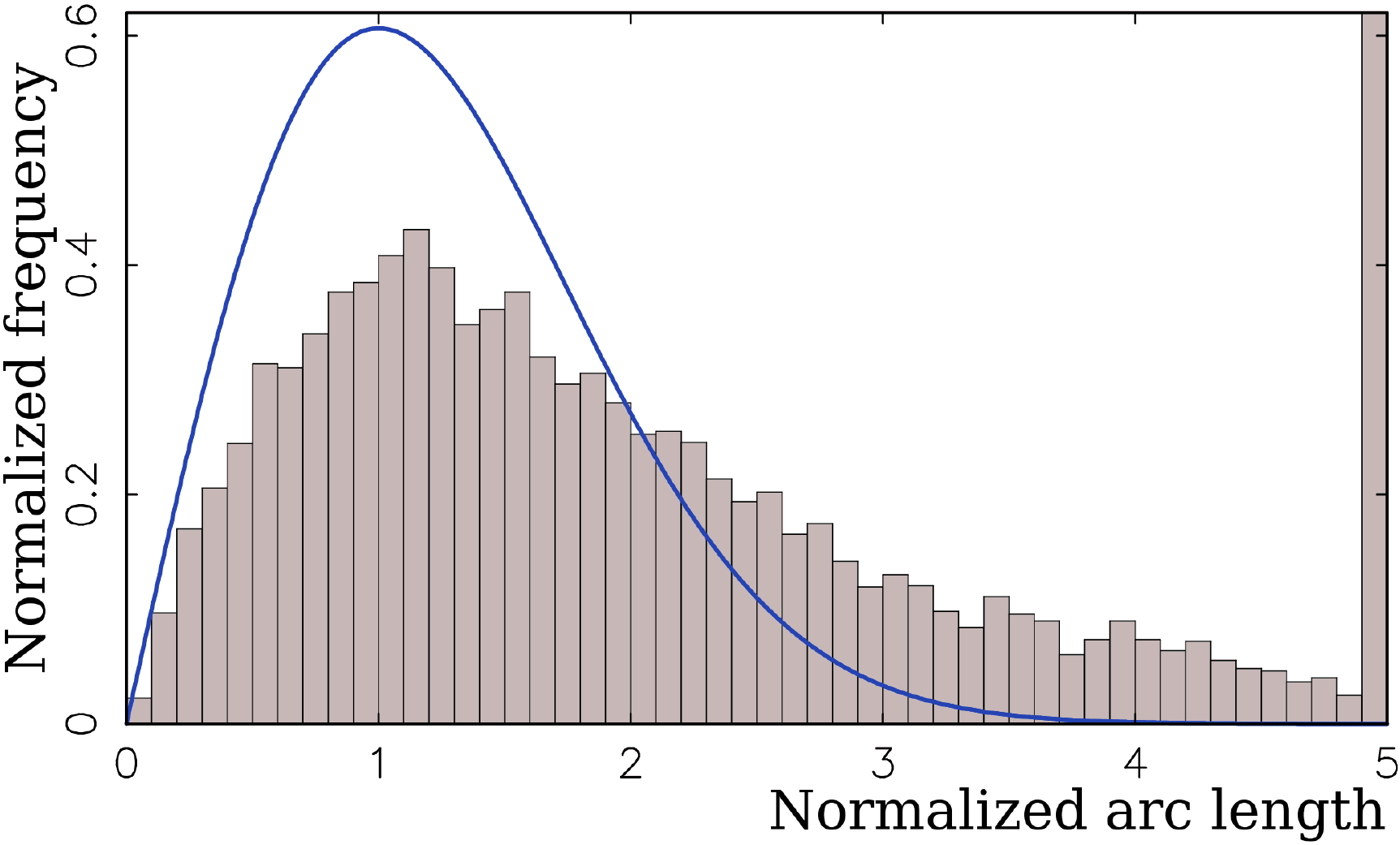}}
    \caption{The distribution of the normalized VLBI/Gaia arc-lengths over 9033
             matched sources. The last bin that holds 1067 normalized arc lengths 
             $>5$ exceeds the plot bounding box. The blue smooth curve shows 
             Rayleigh distribution with $\sigma=1$.
            }
    \label{f:norm_arc_all}
\end{figure}

  Figure~\ref{f:norm_arc_all} shows the distribution of the normalized 
arc-lengths $a/\sigma_a$ among all the matches. The last bin contains 
1067 sources with normalized arcs greater than 5, or 11.4\%. The number of 
sources with normalized arcs greater than 4, what for this work we consider 
statistically significant, is 16.3\%, or 1/6. The goal of our study is 
to explain these outliers.

\begin{figure*}
    \includegraphics[width=0.45\textwidth]{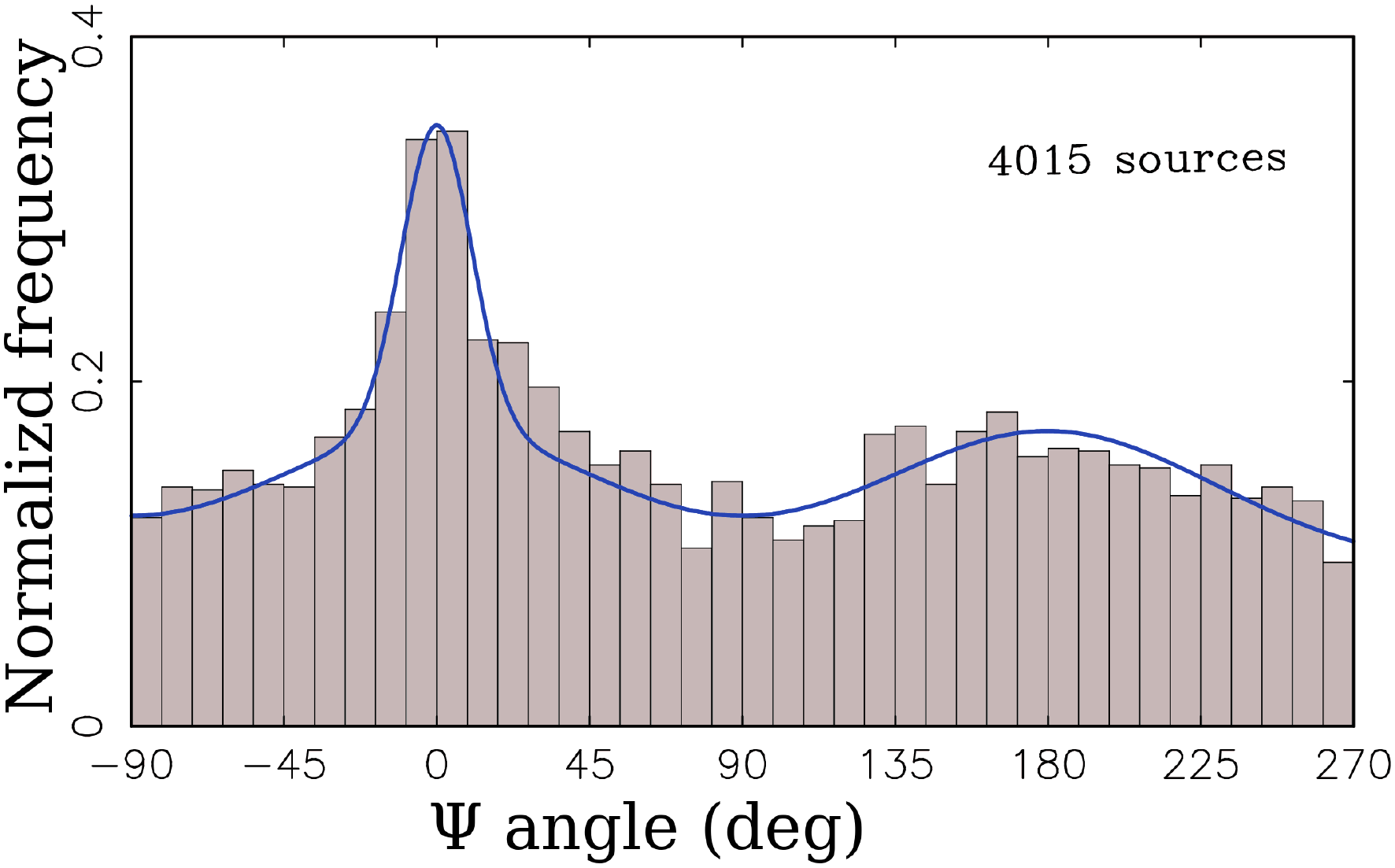} 
    \hfill
    \includegraphics[width=0.457\textwidth]{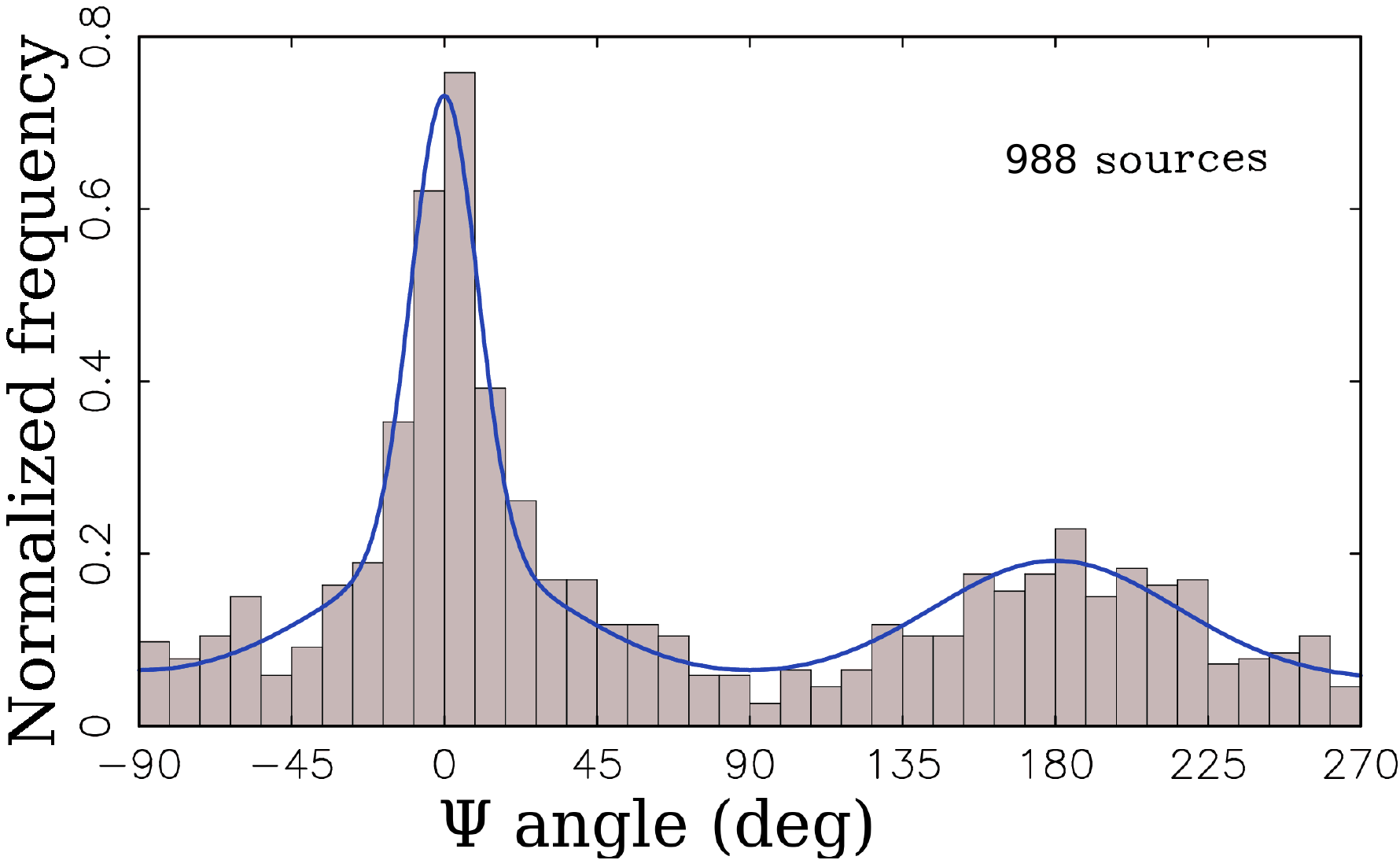}
    \par
    \includegraphics[width=0.45\textwidth]{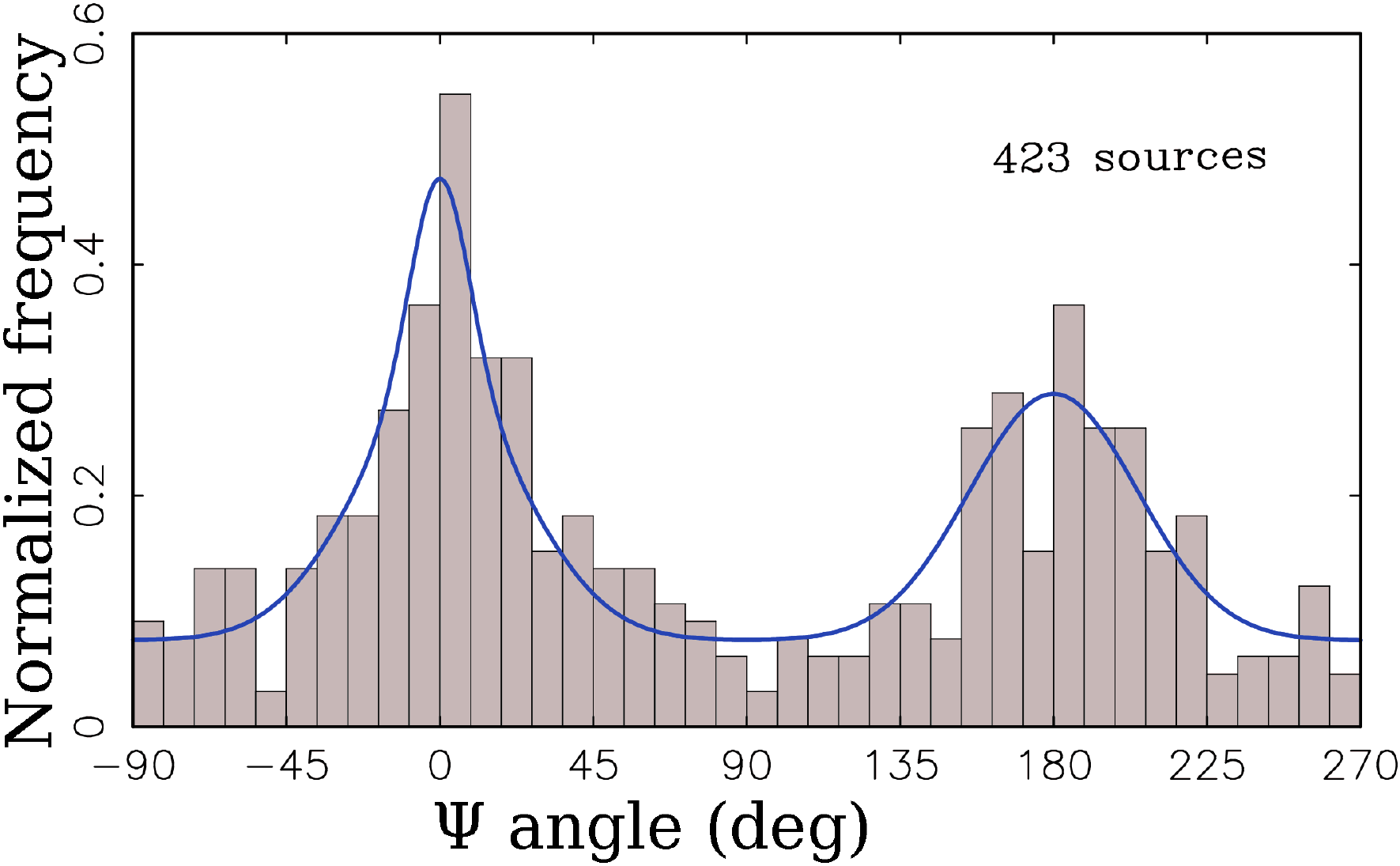} 
    \hfill
    \includegraphics[width=0.45\textwidth]{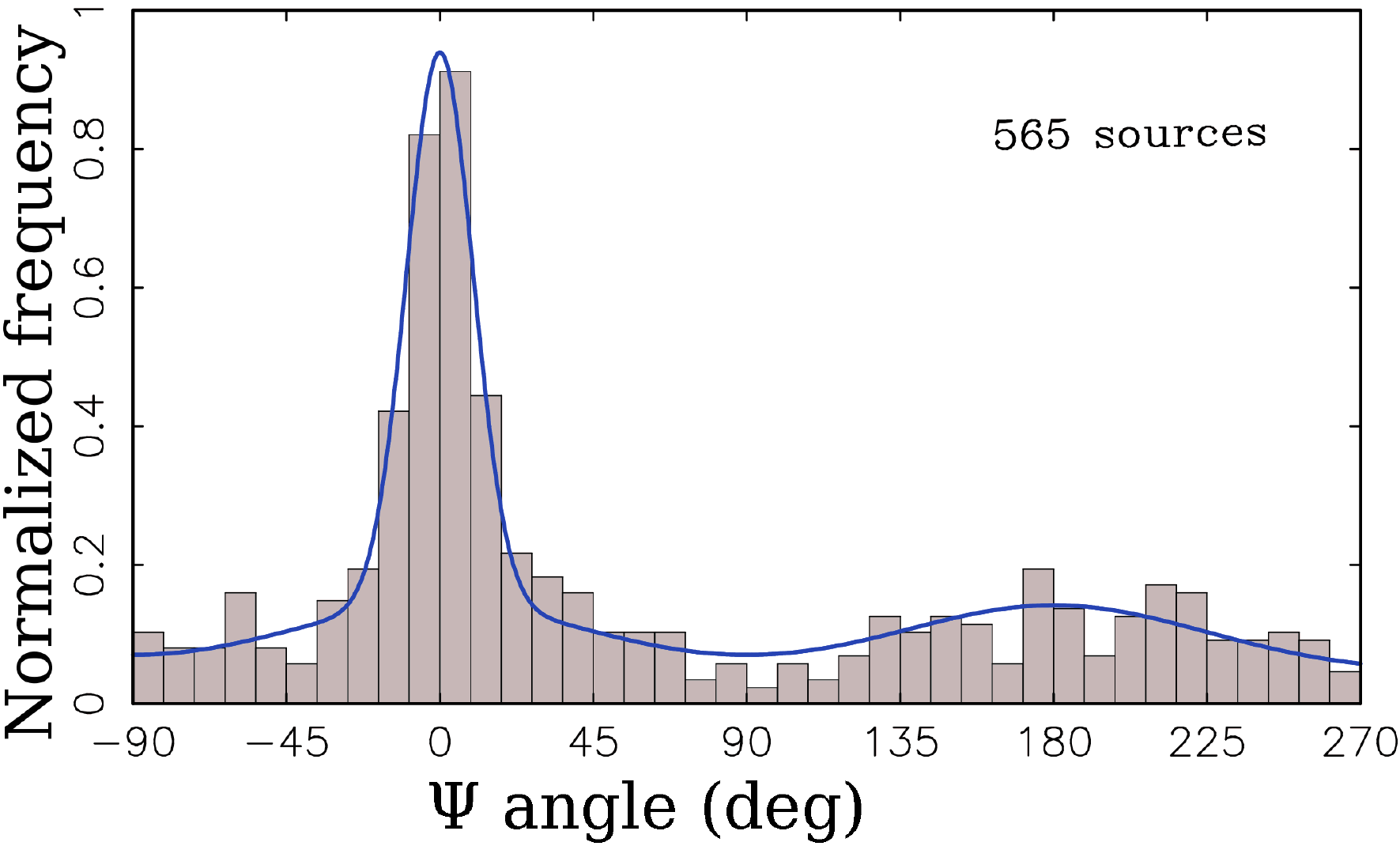}
    \caption{The histograms of the distribution of the position angle
             of Gaia offset with respect to VLBI position counted
             with respect to jet direction counter-clockwise.
             {\it Top left (a):} all the matches with known jet 
             directions. {\it Top right (b):} the matches with 
             $\sigma_\psi < 0.3$~rad. {\it Bottom left (c):} the matches
             with $\sigma_\psi < 0.3$~rad and arc-lengths $<$~2.5~mas.
             {\it Bottom right (d):} the matches with 
             $\sigma_\psi < 0.3$~rad and arc-lengths $>$~2.5~mas.
             Blue curves are the best approximation of a three-component 
             model.
            }
    \label{f:hist_pos}
\end{figure*}

  We computed the histograms of the distribution of the position angle offsets 
with respect to the jet directions determined from the analysis of VLBI images 
at milliarcsecond scales. We denote this quantity as $\psi$. Such a histogram 
is shown in Figure~\ref{f:hist_pos}. Comparing this Figure with the upper left 
Figure~3 in \citet{r:gaia2}, demonstrates that the anisotropy is revealed even 
more clearly: the peaks became sharper and narrower. The height of the peak 
with respect to the background is 2.8 versus 1.7 and the full width at half 
maximum (FWHM) is 0.42~rad versus 0.62~rad. We confirmed that anisotropy of 
$\psi$ angle is not an artifact of Gaia DR1, and the prediction made 
in \citet{r:gaia3} has come true.

  We should note that the histogram of $\psi$ angles is affected by its 
measurement errors that depend on $a/\sigma_\phi$. We assume 
$\sigma_\psi = \sigma_\phi$, neglecting errors in the determination of jet 
direction angles because at the moment, we cannot precisely characterize 
them. At large $a/\sigma_\phi$ (say, more than 4), the distribution of the
$\psi$ errors for a given measurement converges to the normal distribution.
At low $a/\sigma_\phi$ (say less than 0.25), the distribution is converging
to the uniform distribution. The analytic expression for the $\psi$ errors 
can be found in page~233 of \citet{r:tms}. Including measurements of $\psi$ 
with large errors smears the histogram. In order to mitigate smearing, we 
filtered out matches with $\sigma_\psi > 0.3$~rad. We found empirically 
that reducing the threshold further degrades the histograms as a consequence
of the scarcity of remaining points, though does not change their shape 
noticeably.

  Figure~\ref{f:hist_pos}b shows the histogram of $\psi$ angles for all 
the matches with $\sigma_\psi < 0.3$~rad. The peaks at $0^\circ$ and 
$180^\circ$ became much stronger. A further analysis revealed that the 
histograms are different for short and long arc lengths between VLBI 
and Gaia positions as shown in Figure~\ref{f:hist_pos}c 
and~\ref{f:hist_pos}d.

  To characterize the histograms, we fitted a mathematical model to 
them as follows:
\beq
   f(\psi) = \alpha N(0,\sigma_1) \: + \: \beta  N(0,\sigma_2) \: + \: 
             \beta N(\pi,\sigma_2) + \Frac{1 - \alpha - 2 \beta}{2\pi},
\eeq{e:e2}
  where $N(a,\sigma)$ is the normalized Gaussian function with first two
moments $a$ and $\sigma$. We have selected a model that is as simple 
as possible. In the context of this study a choice of functions to 
represent the empirical distribution is irrelevant, as far as the 
mathematical model fits the distribution. Parameter $\alpha$ describes 
the contribution of the main narrow peak, parameter $\beta$ describes 
the contribution of the secondary wide peaks that has the maxima at 
both 0 and $\pi$, and the last term describes the contribution of the 
uniform component of the distribution. We noticed that the broad peaks 
at $\psi=0$ and $\pi$ have a similar shape and fitting them separately 
with two additional parameters does not improve the fit.
The results of fitting this 4-parametric model to the histograms in 
Figures~\ref{f:hist_pos}a--d are shown in Table~\ref{t:hist_fit}.

\begin{table}
   \caption{Results of fitting the model in eq.~\ref{e:e2} to the histograms
            in Figures~\ref{f:hist_pos}a--d.}
   \begin{tabular}{lll@{\:\:}ll@{\:\:\:}lr}
      \hline
      \hspace{-1em} Case & \ntab{c}{$\alpha$} & \sc fwhm${}_1$ & \ntab{c}{$\beta$} & \sc fwhm${}_2$ & $ 1 - \alpha - 2 \beta$ & \# src \\
           &         & rad      &                & rad        &                         &            \\
      \hline
         a & 0.08     & 0.42       & 0.17   & 2.03       & \ntab{c}{ 0.58}     & 4017       \\
         b & 0.23     & 0.40       & 0.22   & 1.48       & \ntab{c}{ 0.33}     &  985       \\
         c & 0.07     & 0.35       & 0.17   & 1.01       & \ntab{c}{ 0.47}     &  423       \\
         d & 0.24     & 0.40       & 0.17   & 1.84       & \ntab{c}{ 0.28}     &  565       \\
      \hline
   \end{tabular}
   \label{t:hist_fit}
\end{table}

   We see that the main peak at $\psi = 0$ with a FWHM around 0.4~rad 
is rather insensitive to the way how a subsample is drawn. We 
tentatively conclude that the fitted FWHM is the intrinsic width of the peak.
The peak at $\psi = 0$ is contributed predominately by the matches with large
position offsets, and it is related to the presence of optical jets.
Several factors contributed to the peak broadening: a)~the intrinsic
jet width; b)~errors in determination of the jet direction;
c)~curvature of the jet, which makes jet direction determination problematic.
Perturbations in the jet shape are magnified because of Doppler boosting. 
Typically, only a beginning of a jet is discernible at VLBI images due to 
limited dynamic range, \Note{while the Gaia centroid is sensitive to jets at 
scales comparable with the PSF and smaller.}

   Two secondary peaks are broad, with maxima at $\psi = 0$ and $\pi$.
They are formed by matches almost exclusively with offsets shorter than 
2--2.5~mas. The fraction of these secondary peaks in the distribution 
weakly depends on the subsample selection, 0.17--0.22, but its FWHM varies
between subsamples. We interpret it as an indication that a simplistic 
4-parameter model is too coarse to fully describe the empirical 
distribution which shape depends on the VLBI/Gaia offset length.

  \Note{The sixth column} in Table~\ref{t:hist_fit} shows the fraction of 
the sources which offset position angles have the uniform distribution, 
i.e., is their offsets are not related to the core-jet morphology. This 
fraction is 0.58 for the histogram made using all the observations. The 
fraction is reduced to 0.33 for the subsample of observations with 
$\sigma_\psi < 0.3$~rad and to 0.25 for the subsample with 
$\sigma_\psi < 0.2$~rad. This reduction occurs partly due to mitigation 
of the histogram smearing, and partly due to the selection bias. Since 
$\sigma_\psi$ depends on both an uncertainty of position estimates and 
an arc-length, selecting a subsample with the upper limit for $\sigma_\psi$ 
disproportionately favours the matches with long VLBI/Gaia offsets that for
given position uncertainties get high chances to have low $\sigma_\psi$.

\begin{figure}
    \centerline{\includegraphics[width=0.45\textwidth]{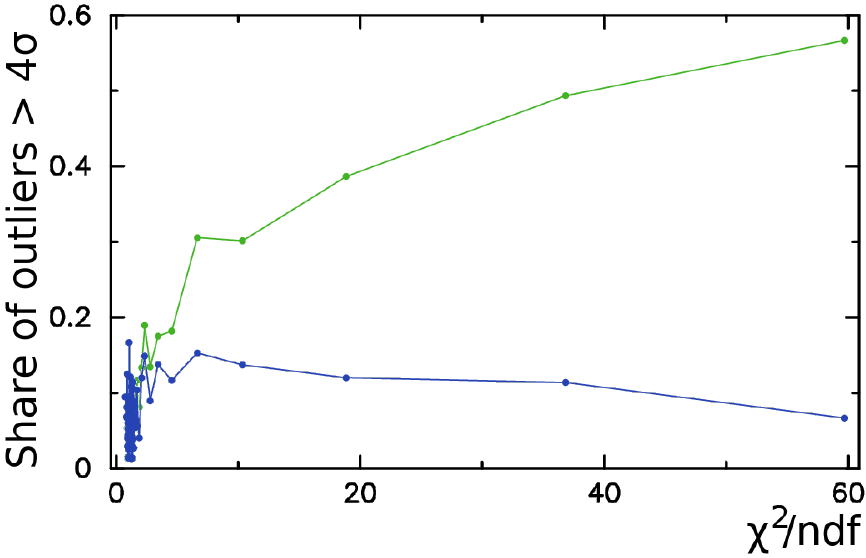}}
    \caption{The fraction of outliers with normalized arc length of VLBI and Gaia 
             matches $>4$ for \note{1\% percentile bins} of $\chi^2$/ndf. 
             The horizontal axis is along the median value of $\chi^2$/ndf within 
             each percentile. The upper green curve was computed using original 
             Gaia position uncertainties. The low blue curve was computed using 
             Gaia uncertainties multiplied by $\sqrt{ \chi^2/\mbox{ndf}}$ factor.
            }
    \label{f:chi_ndf_outliers}
\end{figure}

  The distribution of the VLBI/Gaia position offset angles was studied by 
\citet{r:gaia5} for different purposes applying a different fraction analysis 
approach. The outcome of their study qualitatively agrees with results 
presented here.

\subsection{Re-scaling VLBI and Gaia reported position uncertainties}

  The presence of strong peaks at histograms in Figures~\ref{f:hist_pos} 
means these matches are affected by systematic differences. These differences
also affect the distribution of normalized arc lengths shown in 
Figure~\ref{f:norm_arc_all}. In order to mitigate their impact, we
re-drew the histogram and excluded the sources with $\psi$ angles within 
0.5~rad of peaks at 0 and $\pi$. As a result, we got a clean sample that 
is not affected by the presence of optical jets. We used this clean sample 
for characterizing Gaia and VLBI reported position uncertainties. 
We wanted to answer the question how realistic the uncertainties are.

  We noticed that the number of outliers, i.e., the matches with the 
normalized arc $>4$, grows with an increase of $\chi^2$/ndf, where ndf 
is the number of degrees of freedom. $\chi^2$ is provided in variable 
{\sf astrometric\_chi2\_al} of the Gaia DR2 archive. The number of 
degrees of freedom was computed as the difference of the variables 
{\sf astrometric\_n\_good\_obs\_al} and {\sf astrometric\_params\_solved}.

\Note{We sorted the dataset in the increasing order over $\chi^2$/ndf and 
split it into 100 percentile groups of 91 objects each. Then we computed 
the fraction of outliers for each percentile group.} The dependence of 
the fraction of outliers as a function of the mean $\chi^2$/ndf within 
a percentile is shown with a green curve in 
Figure~\ref{f:chi_ndf_outliers}. It grows approximately as 
$\sqrt{\chi^2/\mbox{ndf}}$ when $\chi^2/\mbox{ndf} > $ 1.5--2. Since 
the number of degrees of freedom is the mathematical expectation of 
$\chi^2$, in a case if all uncertainties of Gaia observables of a given 
source are underestimated by a common factor, multiplying them by 
$\sqrt{ \chi^2/\mbox{ndf}}$ corrects the impact of the measurements error 
underestimation. The blue curve in Figure~\ref{f:chi_ndf_outliers} 
demonstrates that after re-scaling Gaia position uncertainties, the 
dependence of the number of outliers as a function of $\chi^2$/ndf has 
disappeared. Scaling position errors by $\chi^2$/ndf inflates them,
which makes the normalized arc-lengths smaller. We argue that re-scaling 
Gaia position errors makes them more realistic by accounting for 
the additional noise that also increases $\chi^2$/ndf.

  In addition to source-dependent re-scaling that is based on $\chi^2$/ndf
statistics of a given source, we evaluated global scaling factors for
both VLBI and Gaia that affect every source. This is the simplest way to 
mitigate the impact of systematic errors on uncertainties and make them more
realistic without re-running a solution. Since the normalized arc lengths 
are affected by both uncertainties in VLBI and Gaia positions, we estimated 
the scaling factors of VLBI uncertainties by processing the subset of 
observations with $\sigma_{\rm g,maj} > 5 \, \sigma_{\rm v,maj}$
and vice versus: we estimated scaling factors for the Gaia uncertainties 
(after scaling them by $\sqrt{ \chi^2/\mbox{ndf}}\:$)
by processing the subset of observations with 
$\sigma_{\rm g,maj} < 5\, \sigma_{\rm v,maj}$. We adjusted the scaling
factors in such a way that the distribution of normalized arc-lengths
of the subsample be approximated with the Rayleigh distribution $\sigma=1$.
The scaling factors are 1.06 for Gaia and 1.30 for VLBI. Applying scaling 
parameters to uncertainties to account for the contribution of 
systematic errors is a common technique. For instance, a scaling 
factor 1.5~was used to inflate source position uncertainties in the 
ICRF1 catalogue \citep{r:icrf1}.

\begin{figure}
    \centerline{\includegraphics[width=0.45\textwidth]{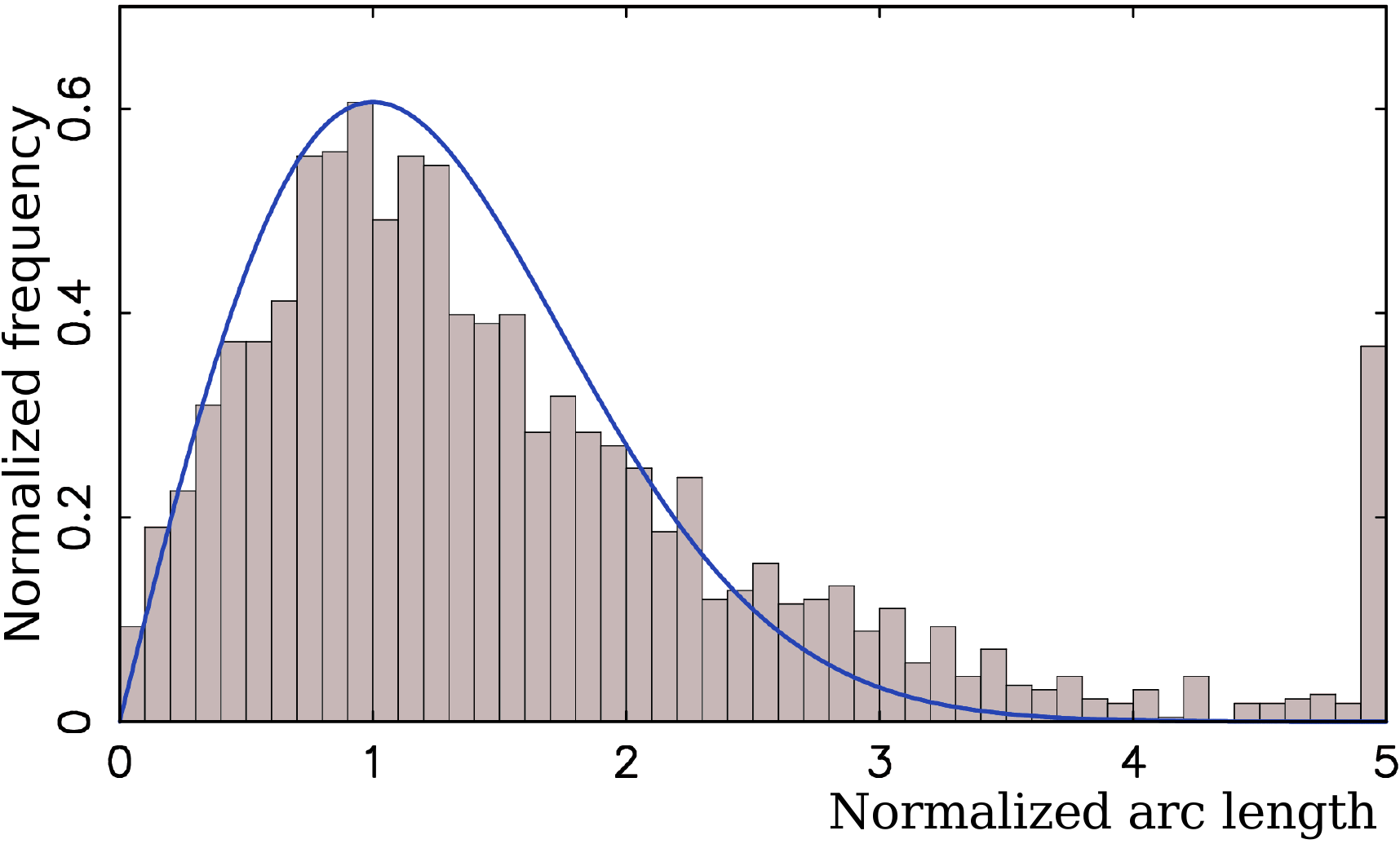}}
    \caption{The distribution of normalized VLBI/Gaia arc-lengths over 2313 
             matched sources. The sample includes all the sources with
             known jet directions and excludes the sources with 
             $\psi \in [-0.5, -0.5]$ and $ \psi \in [\pi-0.5, \pi+0.5]$~rad.
             Scaling factors 1.05 and $1.30 \sqrt{\chi^2/\mbox{ndf}}$ were 
             applied to Gaia and VLBI. The blue smooth curve shows 
             Rayleigh distribution with $\sigma=1$.
            }
    \label{f:norm_arc_off}
\end{figure}

\begin{table}
   \caption{Table with the fraction of matches with normalized residuals
            $>4$ for a number sub-samples in pro cents (column r).
            The last two rows show the sub-samples of matches with 
            known jet directions. The second and fourth raw use 
            a sub-sample of matches with VLBI semi-major error ellipse 
            less than median among all matches and the matches with 
            known jet directions respectively. Column ``off-peak'' 
            excludes the sources with $\psi \in [-0.5, -0.5]$ and 
            $ \psi \in [\pi-0.5, \pi+0.5]$~rad. Column ``on-peak'' 
            include the sources with $\psi$ in these ranges and 
            exclude everything else. 
           } 
   \begin{tabular}{l @{\enskip\:\:} r@{\:\:}r r@{\:\:}r r@{\:\:}r}
   \hline
         & \nntab{c}{all} & \nntab{c}{off-peak} & \nntab{c}{on-peak} \\
         &   \ntab{c}{r} & \# src & \ntab{c}{r}& \# src & \ntab{c}{r} &  \# src    \\
   \hline
                               all &  9.0  & 9033    &  6.6 & 7288 & 19.4 & 1702 \\ 
   $ \sigma_v \le 0.963$ mas       & 10.0  & 4496    &  5.9 & 3169 & 19.7 & 1323 \\ 
   \hline
    all with known $\psi$          & 11.2  & 4017    &  5.4 & 2313 & 22.1 & 1702 \\ 
   $ \sigma_v \le 0.455$ mas       & 11.4  & 1997    &  4.3 & 1109 & 20.3 &  888 \\ 
   \hline
   \end{tabular}
   \label{t:share_outliers}
\end{table}

  Since as we have established, the Gaia systematic errors in AGN positions
caused by optical structure have a strong concentration towards $\psi=0$ and 
$\psi=\pi$, we expected that the removal of the matches with 
$\psi \in [-0.5, -0.5]$ and $ \psi \in [\pi-0.5, \pi+0.5]$~rad and keeping 
only ``off-peak'' matches should affect the statistics of the number of 
outliers. We computed the fraction of matches with normalized residuals 
$>4$ for for several sub-samples. Since we applied error re-scaling, the 
number of outliers has reduced with respect to our initial estimate 
mentioned above. The first row of Table~\ref{t:share_outliers} shows 
that excluding the sources within the peaks of the distribution of $\psi$ 
angle reduces the number of outliers by a factor of 1.36. On the contrary,
considering only the sources within 0.5~rad of the peaks doubles the number 
of outliers. Since the jet directions were determined only for 45\% of the 
matches, these statistics underestimate the impact of the presence of 
optical jets on Gaia positions. If to count only the sources with known jet 
directions, excluding the sources within the peaks reduces the number of 
outliers by a factor of 2.07. Rows 2 and 4 of Table~\ref{t:share_outliers} 
shows also the statistics for the subsamples of low 50\% percentile of 
VLBI re-scaled errors. The reduction of the number of outliers is 1.77 for 
the 50\% percentile of the overall sample of matched sources and 2.65 for 
the sub-sample of the sources with known jet directions. The reduction of the
number of outliers is greater for the lower 50\% percentile because 
the sources with smaller position uncertainties have smaller errors in 
determining the $\psi$ angle, what makes discrimination of the ``on-peak'' 
and ``off-peak'' sources more reliable.

  Results in Table~\ref{t:share_outliers} show that the presence
of optical structure parallel to the jet explains 62\% of VLBI/Gaia position
offsets significant at the $4 \sigma$ level for a sub-sample of 23\% of
VLBI/Gaia matches that have known jet directions and VLBI position errors
lower than the median. In order to generalize this result to the entire 
population of radio-loud AGNs, we need assume that the significance of 
VLBI/Gaia offsets does not depend on VLBI position error and does not 
depend on the measurability of the radio jet directions. The VLBI position 
errors above 0.2--0.3~mas level are limited by the thermal noise, and thus, 
the first assumption is valid. The validity of the second assumption is 
questionable. The detectability of parsec-scale radio jet depends on the 
jet brightness and the dynamic range of observations that in turn depends 
on the source flux density. Since the correlation between radio and optical 
fluxes is low, missing a jet just because a source was weak does not 
create a selection bias. However, if a jet direction for a given source 
was not detected because its radio jet is intrinsically weaker, missing 
such a source may create a selection bias, because a weak radio jet may 
imply a weak optical jet. A sub-sample of sources with determined jet 
direction may have a selection bias towards jets brighter in radio and 
optic with respect to the overall population.

\subsection{The impact of systematic errors on determination of the 
            orientation of the Gaia catalogue with respect to the VLBI 
            catalogue}

  Any source catalogue can be rotated at an arbitrary angle, and the 
observables, e.g., group delays, remain the same. The orientation of
a catalogue can be described by three angles. These three angles cannot 
be determined from observations in principle and are {\it set} 
by imposing certain conditions. The orientation of the RFC catalogue is
set to require the net rotation with respect to the 212 so-called ``defining''
sources in the ICRF1 catalogue \citep{r:icrf1} be zero. \Note{The orientation 
of Gaia DR2 catalogue was established to have zero rotation with to respect 
to 2843 counterparts in the ICRF3-prototype catalogue using the frame 
rotator technique described in detail in \citet{r:agis}}. 

   The systematic differences caused by the optical structure affect the 
procedure for establishing the catalogue orientation. To provide
a quantitative measure of sensitivity of the orientation angles to 
systematic errors, we computed the three angles of Gaia DR2 orientation with 
respect to the RFC VLBI catalogue (See Table~\ref{t:rot_ang}). We see that
selecting different samples, including those the most affected by 
systematic errors (on-peak) and least affected (off-peak), resulted in 
differences in orientation angles around 0.02~mas. A large value of 
the orientation angle around axis 2 is somewhat unexpected, but since 
the ICRF3-prototype catalogue used for alignment of the Gaia DR2 is 
not publicly available, the origin of this relatively large value cannot 
be established.

\begin{table}
   \caption{Estimates of rotation angles around axes 1,2,3 of the Gaia positions of 
            matches with respect to VLBI positions of four sub-samples. Units are 
            milliarcseconds.
           }
   \small
   \begin{tabular}{@{\hspace{-0.2em}} l @{\quad}r@{\quad} r @{\enskip\enskip} r @{\enskip\enskip} r}
      \hline
                 & \# Obs & \ntab{c}{Axis 1}   & \ntab{c}{Axis 2}     & \ntab{c}{Axis 3 }    \\
      \hline
       all       & 9033 & $ -0.030 \pm 0.004 $ & $  0.090 \pm 0.004 $ & $ -0.030 \pm 0.005 $ \\
       with jets & 4016 & $ -0.010 \pm 0.005 $ & $  0.092 \pm 0.005 $ & $ -0.010 \pm 0.006 $ \\
       off-peak  & 2647 & $ -0.013 \pm 0.006 $ & $  0.095 \pm 0.006 $ & $  0.008 \pm 0.007 $ \\
       on-peak   & 1369 & $ -0.005 \pm 0.008 $ & $  0.091 \pm 0.007 $ & $ -0.037 \pm 0.009 $ \\
      \hline
   \end{tabular}
   \label{t:rot_ang}
\end{table}

\section{Analysis of Gaia and VLBI proper motions}

\begin{figure}
    \includegraphics[width=0.45\textwidth]{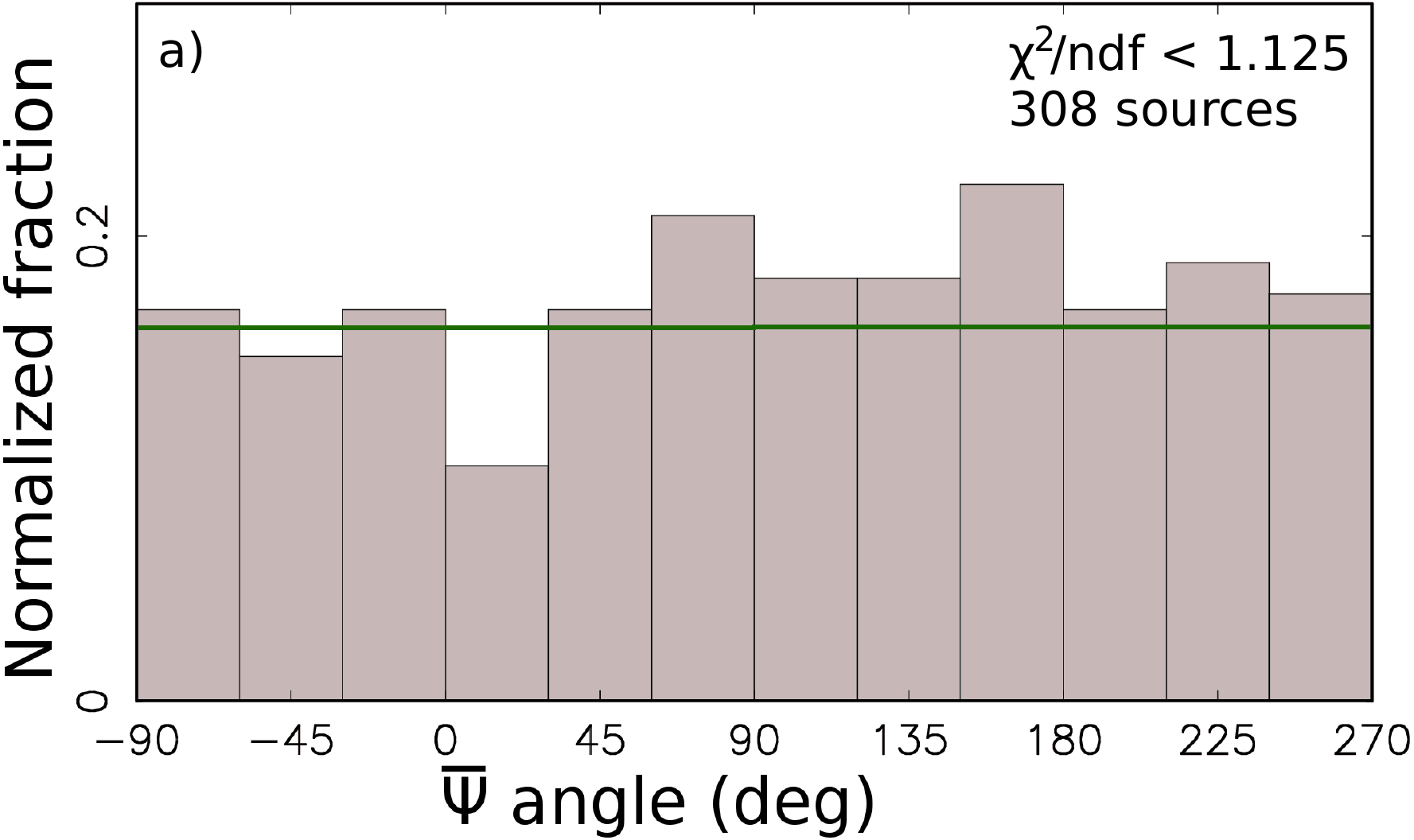}
    \par\smallskip\par
    \includegraphics[width=0.45\textwidth]{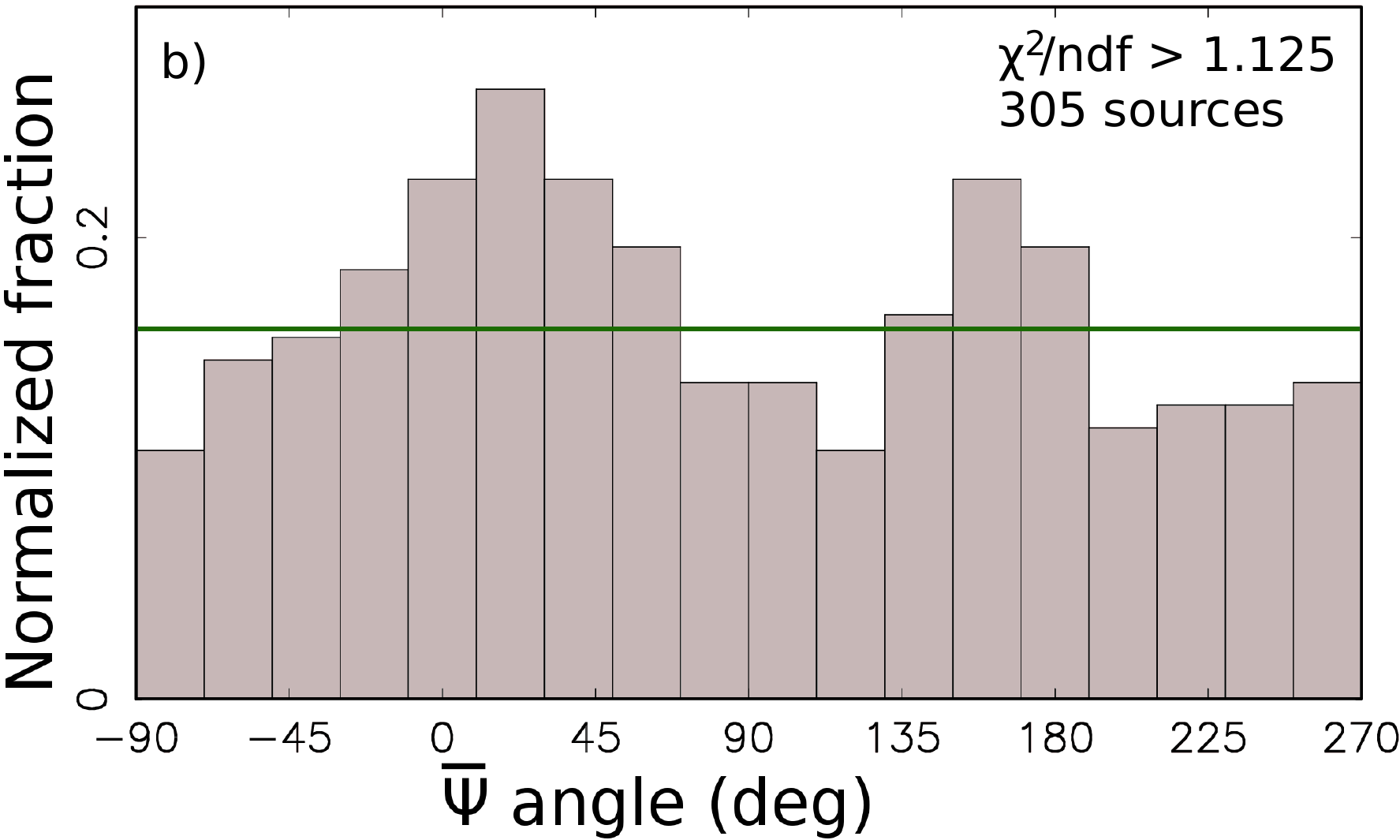}
    \caption{The histograms of the distribution of Gaia proper motion
             position angle with errors $< 0.4$~rad among the matches
             with different $\chi^2$/ndf. {\it Up figure} uses
             the matches with $\chi^2$/ndf less than the median in 
             this subsample 1.125. {\it Low figure} uses
             the matches with $\chi^2$/ndf greater than the median in 
             this subsample 1.125. For comparison, green line shows 
             the uniform distribution.
            }
    \label{f:hist_chi_pm}
\end{figure}

   The Gaia DR2 provides proper motions and parallaxes for 78\% sources.
Among 9081 matches, proper motion estimates are available for 7774 sources.
Since the AGNs are located at cosmological distances, their proper motions 
considered as a bulk tangential displacements are supposed to be well below 
the Gaia detection limit. A flare at the accretion disk or jet will change
position of the centroid. A flare will cause a shift in the position of the 
centroid, and therefore, will result in a non-zero estimate of proper motion.
Such a proper motion may be statistically significant even at Gaia level 
of accuracy. To check it, we made histograms of proper motions as 
a function of the position angles of the proper motion with respect to the 
jet directions denoted as $\bar{\psi}$. We analyzed the sample of 613 
matched sources with $\sigma(\bar{\psi}) < 0.4$~rad. The histograms 
showed weak peaks. The peaks become much sharper when we split the sample 
into two subsets: the subset with $\chi^2$/ndf less then the median 
1.125 and the subset with $\chi^2$/ndf greater than the median 
(See~Figure~\ref{f:hist_chi_pm}).

  We see that the subsample of matches with large $\chi^2$/ndf shows
two peaks at $\bar{\psi}=0$ and $\bar{\psi}=\pi$ that are significant, 
while the subsample with $\chi^2$/ndf below the median does not.
A non-linear motion is one of the reasons why $\chi^2$/ndf deviates
from 1. The histogram at Figure~\ref{f:hist_chi_pm}b tells us that 
among the sources with non-linear motion, the fraction of objects with
proper motions \Note{along or opposite to jet direction} is 
disproportionately high. This dependence on angle $\bar{\psi}$ implies 
that the proper motion is caused by the photocenter changes parallel 
to the jet direction at least for a fraction of the sources.

\begin{figure*}
    \includegraphics[width=0.32\textwidth]{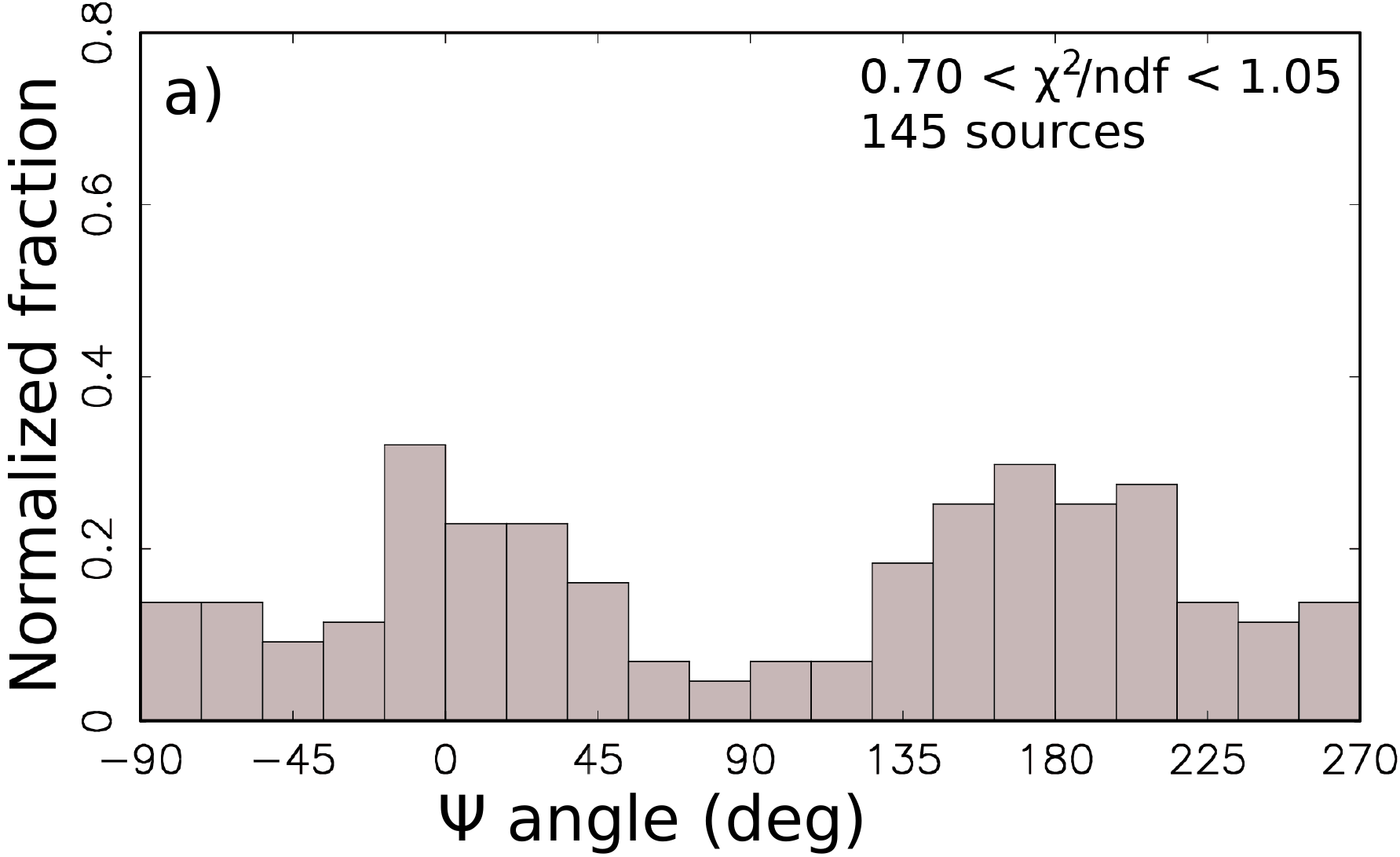}
    \hspace{0.002\textwidth}
    \includegraphics[width=0.32\textwidth]{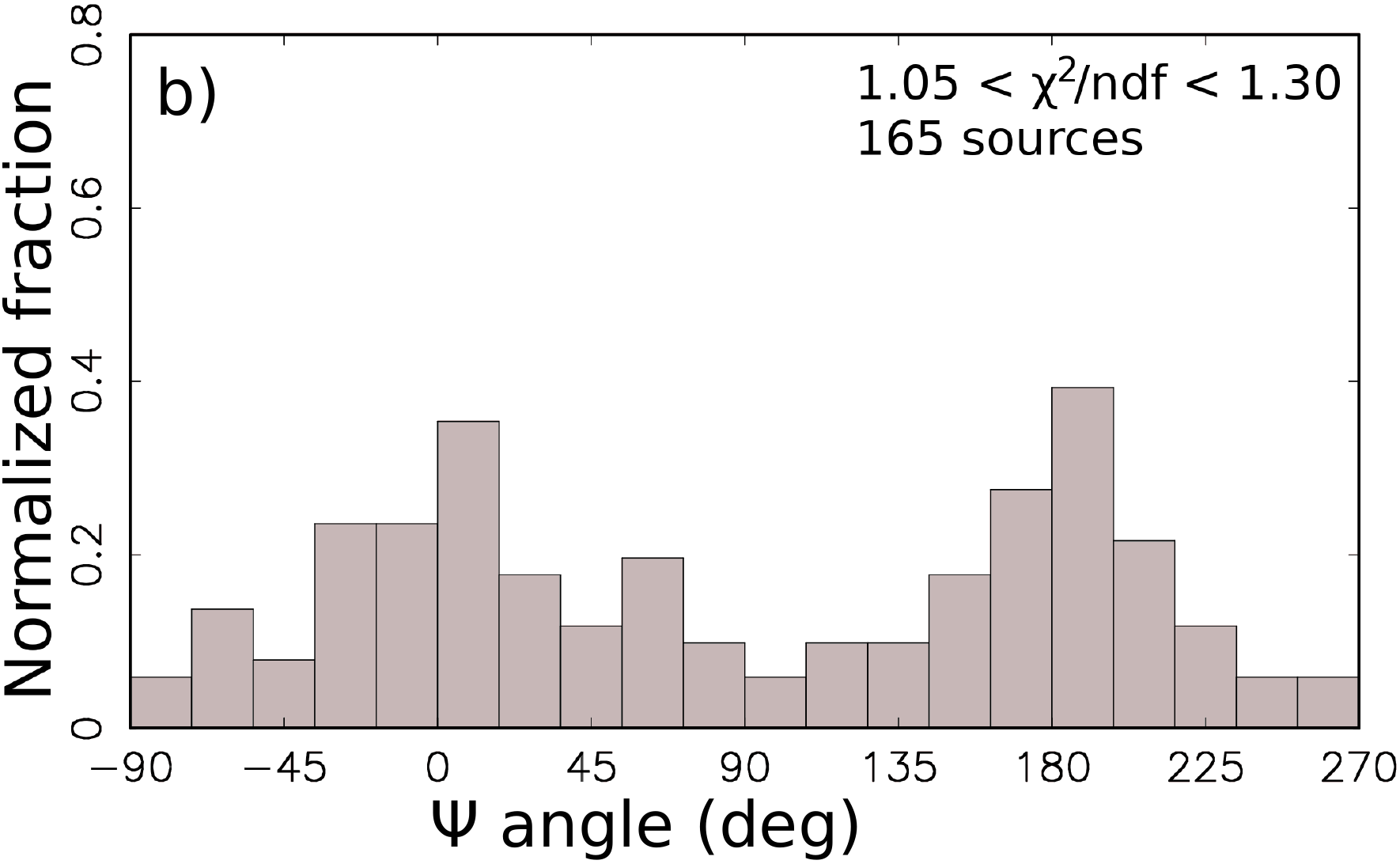}
    \hspace{0.002\textwidth}
    \includegraphics[width=0.32\textwidth]{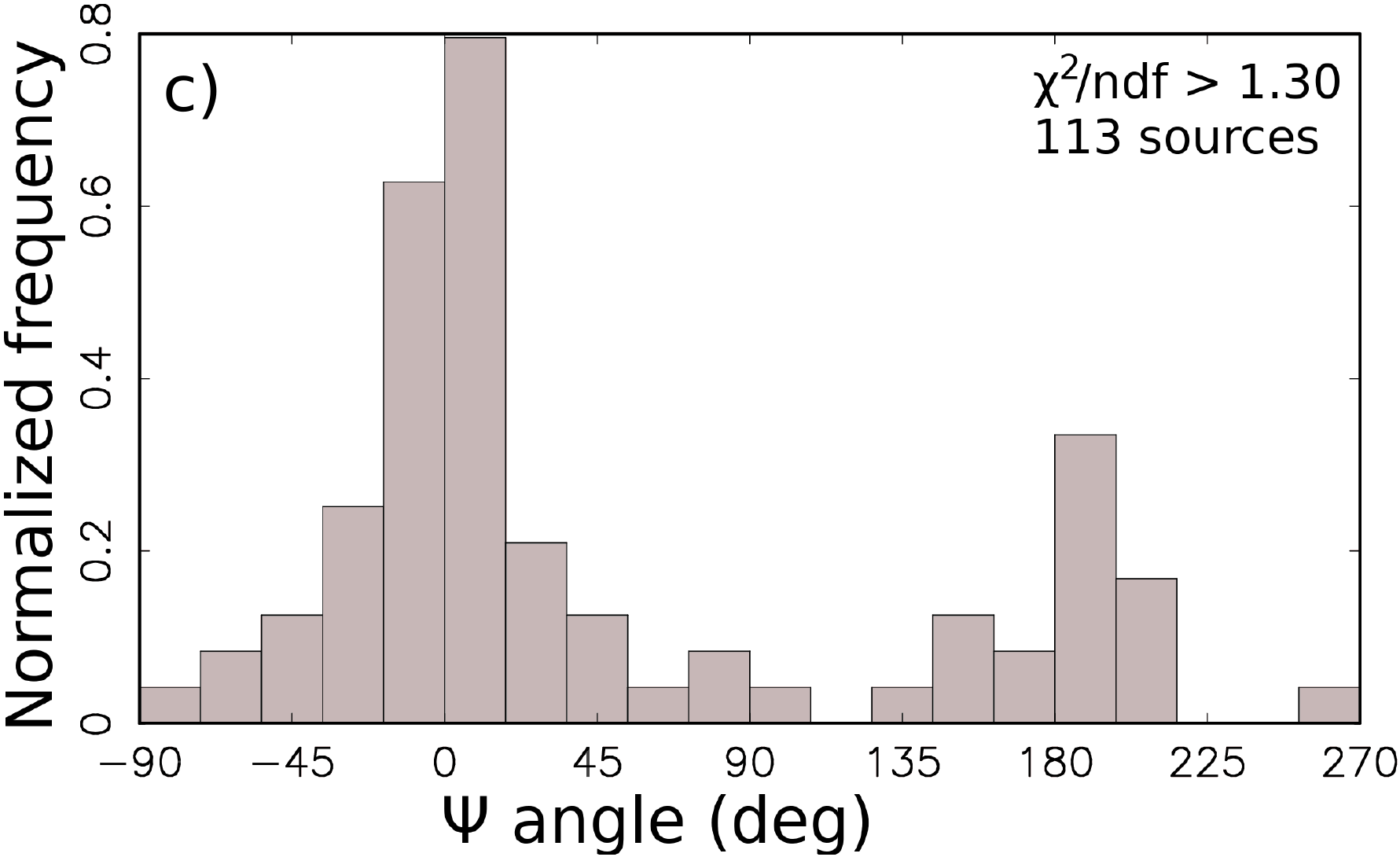}
    \caption{The histograms of the distribution of the position angle
             of Gaia offsets with respect to VLBI positions for matches 
             with $\sigma_\psi < 0.3$~rad and arc-lengths $<$~2.5~mas
             and different ranges of $\chi^2$/ndf.
            }
    \label{f:hist_chi_pos}
\end{figure*}

  We expect that most optical flares happen close to the center of an
AGN, either in the accretion disk or in the jet base. We can not directly 
see where the optical jet flares occur. However, the following arguments 
apply. The radio variability is associated with the apparent jet base --- 
the core \citep[e.g.,][]{r:mojave4,r:mojave13}. Optical synchrotron emission 
is more transparent with even brighter core and steeper jet spectrum 
\citep[e.g.,][]{r:mimica09}. As a result, the jet base is expected to be the 
prime source of optical flares. The correlation between direction of 
linear polarization between optical flares and radio core reported by 
\citet{r:bu07} confirms this.

  Brightening a jet component shifts the centroid temporarily and 
irregularly. We call this behavior jitter and we predicted it 
in \citet{r:gaia3}. Unlike to proper motions of stars, extending 
the observation interval does not result in a convergence of a proper 
motion estimate to some value with small uncertainty. Instead, it slowly
converges to zero. Peaks at $0$ and $\pi$ in the histogram of 
$\bar{\psi}$ over the sub-samples with high $\chi^2$/ndf provide us 
the first evidence that predicted jitter indeed takes place. We used here 
estimates of AGN proper motions and $\chi^2$/ndf as a proxy for 
jitter detection.

  We explored further the impact of a selection based on $\chi^2$/ndf 
on the distribution of position offset angles with respect  to jet 
direction. We did not find a noticeable impact of $\chi^2$/ndf for 
VLBI/Gaia offsets longer 2.5~mas, but we found such a selection affects
the matches with VLBI/Gaia offsets shorter 2.5~mas. 
Figure~\ref{f:hist_chi_pos} shows the distributions of $\psi$ angles
for matches with $\sigma(\psi)<0.3$~rad divided into three sub-samples
approximately equally distributed over $\chi^2$/ndf. The peaks at 
$\psi=0$ and $\psi=\pi$ are broad for the sub-sample of low 
$\chi^2$/ndf. They are getting sharper for the sub-sample of 
intermediate $\chi^2$/ndf. The sub-sample with large $\chi^2$/ndf
is strikingly different than the sub-sample with low $\chi^2$/ndf:
the histogram has a very strong peak at $\psi=0$, i.e., along the jet
direction, and a smaller fraction of matches outside the main peaks.

\begin{figure}
    \includegraphics[width=0.45\textwidth]{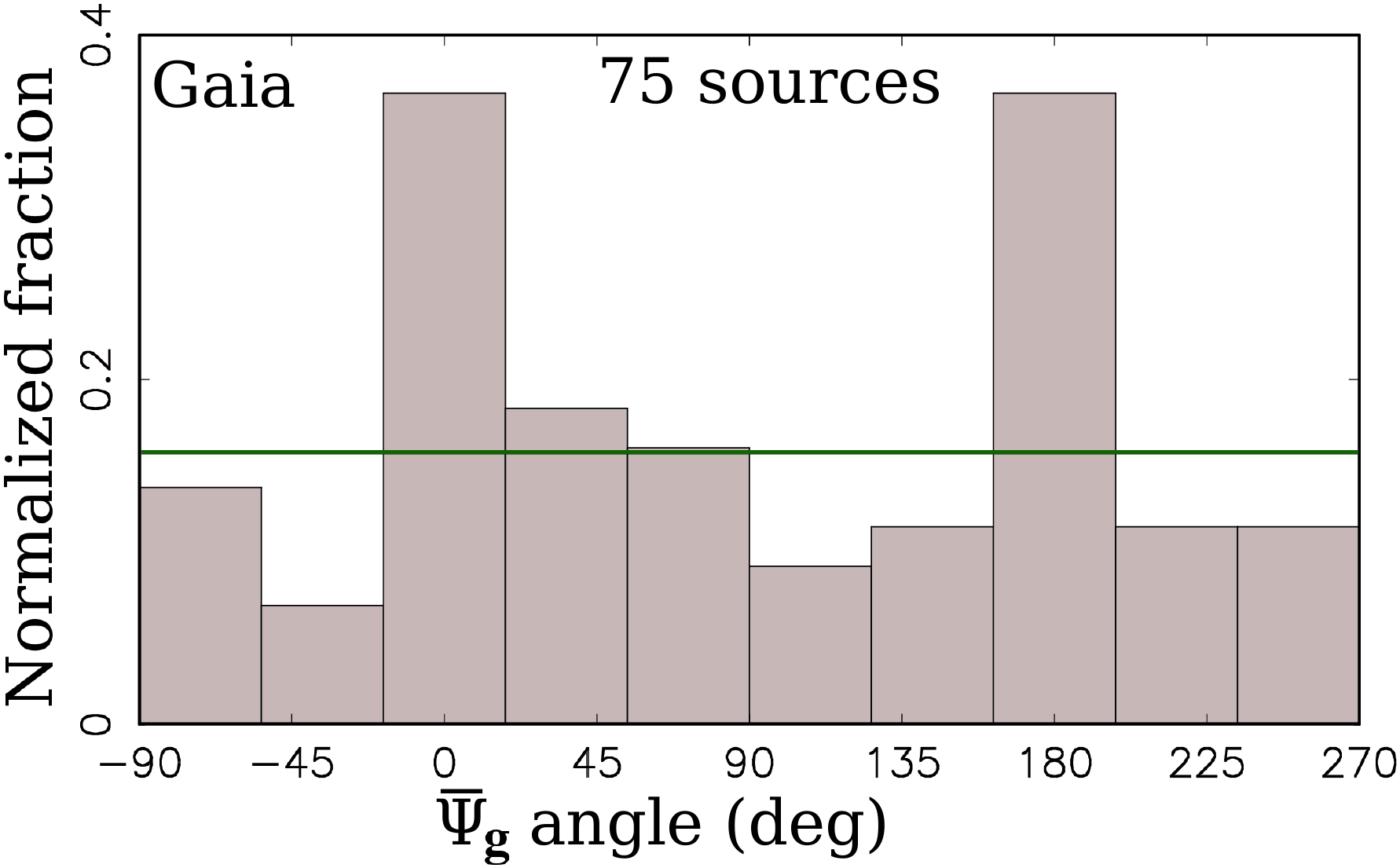}
    \par\smallskip\par
    \includegraphics[width=0.45\textwidth]{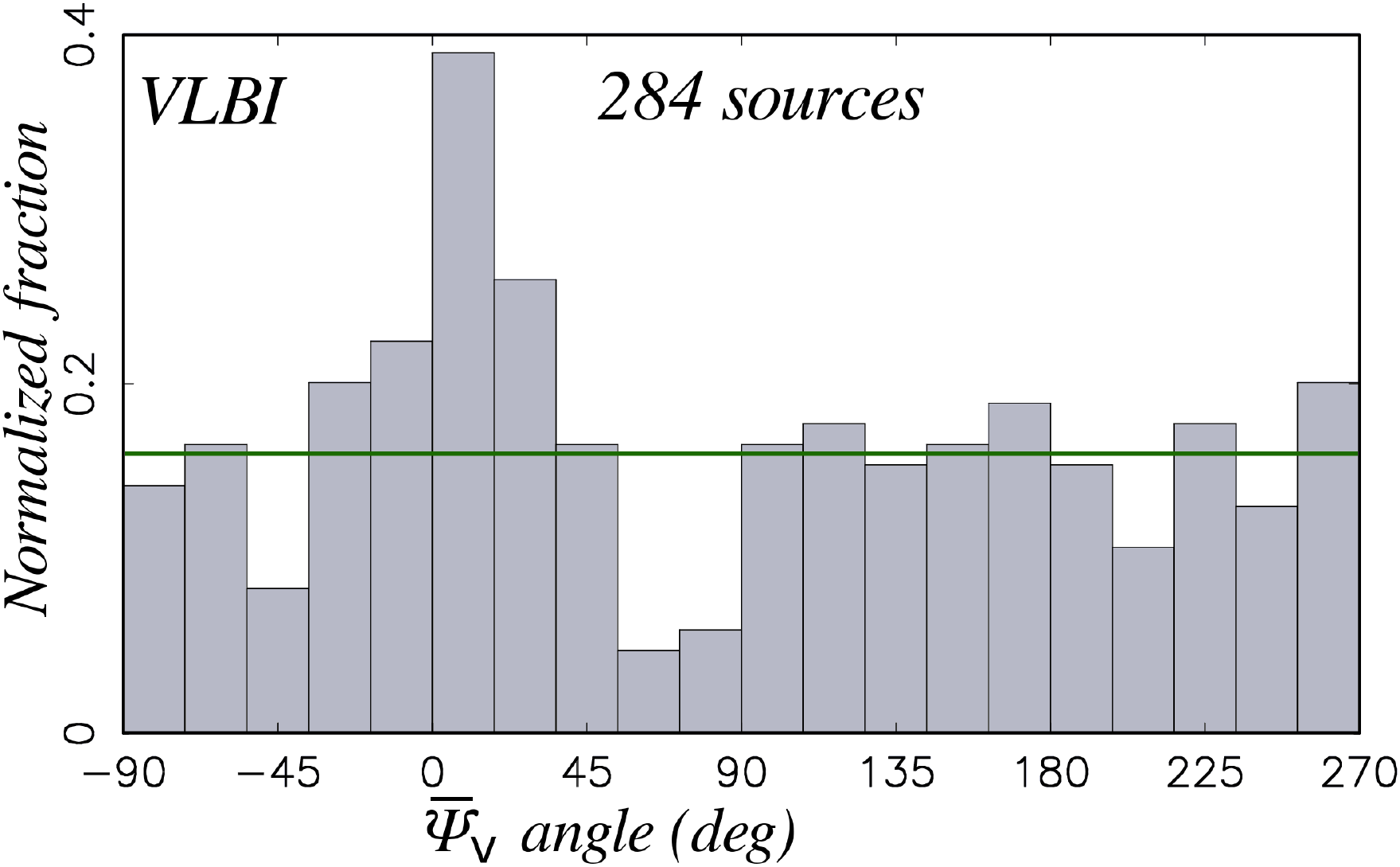}
    \caption{The histograms of significant AGNs proper motion position
             angles with respect to jet directions among matched 
             sources $\bar{\Psi}_{\rm g}$ for Gaia and 
             $\bar{\Psi}_{\rm v}$ for VLBI. {\it Up: } the Gaia DR2 
             proper motions with magnitudes $>3\sigma$ in both proper 
             motions and position offsets. 
             {\it Down: } the proper motions from 
             the VLBI global solution. The horizontal green line shows 
             the uniform distribution.
            }
    \label{f:hist_pm}
\end{figure}

  Analysis of the connection of the Gaia DR2 proper motions with 
$\chi^2$/ndf suggests that the matches with large $\chi^2$/ndf are 
more prone to exhibit the jitter. This allows us to conclude tentatively 
that among the sub-sample of sources with VLBI/Gaia offsets shorter than 
2.5~mas, flares and jitter occur predominantly in the objects 
that have Gaia offsets along the jet direction. This indicates that the 
mechanism that causes an increase of $\chi^2$/ndf may not work or
at least is not dominating for sources with $\psi=\pi$. At the same 
time, Figures~\ref{f:hist_chi_pm} and \ref{f:hist_pm} suggest there is 
no strong preferable sign of the motion direction, either along or 
opposite to the jet. Such a pattern is consistent with a jitter caused 
by flares: depending on when a flare has happened, at the beginning or 
the end of observing interval, the direction of the proper motion 
may be opposite.

  It is instructive to examine whether proper motions in AGN positions 
derived from VLBI data analysis show the same pattern. \Note{We ran
a special VLBI solution using all available ionosphere-free linear 
combinations of group delays at 8.4 and 2.3~GHz since 1980 through 
01 August 2018 and estimated proper motions of 3039 sources using 
least squares.} Source structure was considered as a $\delta$-function in
processing VLBI observations during both fringe fitting and computation of
theoretical group delays. We selected the sources that were observed in 
at least 2 sessions over an interval of at least 3 years and each observing 
session had at least 20 usable combinations of group delays.
We applied the data reduction for the acceleration of the barycenter 
of the Solar system towards the Galactic center with right 
ascension 17h45m36.6s, declination $-28^\circ56'00''.0$, and 
magnitude $1.845 \cdot 10^{-10}$~m/s${}^2$. We applied no-net-rotation
constraints on proper motion estimates among 628 sources with strong 
history of observations, namely, observed in at least 8 sessions over 
4 years or longer and have at least 128 usable linear combinations of 
group delays.

  Figure~\ref{f:hist_pm} shows the histograms of the proper motion
position angles $\bar{\psi}_{\rm g}$ and $\bar{\psi}_{\rm v}$ with respect 
to jet directions among those matched sources from Gaia and VLBI that 
have magnitudes of the proper motions and position offsets significant 
at $3\sigma$ level for Gaia and $4\sigma$ for VLBI. There are 75 such sources
in Gaia dataset and 284 in the VLBI dataset. The fraction of Gaia sources
in bins at $\bar{\psi}_g=0$ and $\bar{\psi}_g=\pi$ is a factor
of 3 greater than on average. The median proper motions in these 
samples is 1.15~mas in the Gaia subset and 0.022~mas in the VLBI 
subset, i.e., a factor of 52 less. The Gaia proper motions were evaluated
over the 1.15~year time interval. The VLBI proper motions were evaluated
over a time span in the range of 7.9 to 38.2~years with the median 
26.5~years, a factor of 22.8 longer. The median magnitude of proper motions 
parallel to jet directions does not differ from the median magnitudes
over the entire populations for both VLBI and Gaia.

  While the histogram of the Gaia proper motion position angles shows
peaks at both $\bar{\psi}_g=0$ and $\bar{\psi}_g=\pi$, a similar 
histogram of the VLBI proper motion position angles shows only
a peak at $\bar{\psi}_v=0$. Explanation of this pattern in VLBI 
proper motions requires further investigation. As we showed in
\citet{r:gaia3}, unlike to a power detector, e.g. a CCD, an interferometer 
is not sensitive to the centroid change. Unaccounted contribution of an 
extended jet affects source position estimates at scales of tens 
microarcseconds. The unaccounted contribution of source structure to VLBI 
positions may reach a level of 0.1--1~mas if the image has more than one 
compact component, especially if the compact component is located at 
a distance comparable with a resolution of the interferometer. A change in
the relative brightness or the distance between components due to flares 
causes a change in position estimates at given epochs, and as result, proper
motion. The peak \Note{around $\bar{\Psi_v}=0$ at the low plot 
in Figure~\ref{f:hist_pm} confirms that at least for some sources this 
mechanisms works.}

\section{Other known causes of VLBI/Gaia positions offsets}

  A number of authors \citep{r:gaia_icrf2,r:mak17,r:Frouard18}
suggested alternative explanations of statistically significant offsets:

\begin{itemize}
   \item Error in matching VLBI and Gaia objects. They are easily controlled 
         by computing the probability of false association based on the source 
         density in the vicinity of the candidates to association. The 
         cutoff of the probability of false association $2 \cdot 10^{-4}$ 
         results in the mathematical expectation of the total number of false 
         associations to 2. The coarseness of the source density model may 
         increase the number of false associations, but very unlikely it can 
         increase their count by an order of magnitude. 

   \item An extended galaxy around a quasar. Position estimates of 
         extended objects may suffer from deficiencies of the current Gaia 
         PSF model. To examine to  which extent this affected VLBI/Gaia 
         offsets, we investigated a subsample of the galaxies from 
         the NGC catalogue \citep{r:ngc}. \Note{Since these are the brightest
         known background galaxies, if emission of background galaxies affects
         VLBI/Gaia offsets, the contribution of such emission 
         is supposed to be the highest among the NGC subsample.} We used 
         positions of these sources from Simbad database \citep{r:simbad}, 
         cross-matched them against the RFC catalogue, and found 167 associations. Of them, 49, or 29\%, 
         have a counterpart in Gaia DR2. It is worth noting that the fraction 
         of VLBI/Gaia matches \Note{among NGC galaxies} is twice less than in 
         the full sample. Without re-scaling the Gaia position uncertainties 
         by the $\sqrt{\chi^2/\mbox{ndf}}$ factor, approximately one half of 
         these counterparts, 26 objects, have normalized arc-lengths exceeding~4.
         However, these objects have large $\chi^2/\mbox{ndf}$. After 
         re-scaling the Gaia position uncertainties, all but one objects
         had the normalized arc-length below~4.

            We conclude that extended galaxies may have large VLBI/Gaia offsets,
         but they also have large $\chi^2/\mbox{ndf}$. Scaling the uncertainties
         by the $\sqrt{\chi^2/\mbox{ndf}}$ makes the normalized arc-lengths
         of galaxies indistinguishable from the rest of the sample.
          
   \item Lensed quasars. There are 10 known gravitational lenses in the sample
         of VLBI/Gaia matches. Since gravitational lenses were extensively
         hunted using radio surveys \citep[e.g.,][]{r:class}, it is unlikely that
         the RFC has more than several missed gravitational lenses.

   \item Double quasars. \citet{r:mak17} presented a list of 28 sources with 
         VLBI/Gaia DR1 significant offsets that have a close component on 
         PanSTARRS images. Of them, 24 were found in Gaia DR2
         and passed our test of the probability of false association 
         $2 \cdot 10^{-4}$. Of them, 11 have significant VLBI/Gaia DR2 
         offsets. The second component may be either a star or a merging 
         galaxy. During galaxy mergers, the nuclei may be dislodged with 
         respect \Note{to the center of mass of each individual galaxy}. 
         A study of such systems may help to constraint theories of galaxy 
         mergers. However, the number of such systems is small (11 out of 
         2293 identified in \citet{r:mak17}, i.e., 0.5\%).
\end{itemize}

\section{Summary and conclusions}

   Here we summarize the main results of our comparison of AGN 
positions and proper motions from the Gaia DR2 against the most complete 
catalogue of VLBI positions to date, the RFC.

\begin{enumerate}
   \item The Gaia DR2 AGN position uncertainties of VLBI matched sources 
         are a factor of two smaller than the VLBI position uncertainties.
         \Note{Gaia position catalogues are becoming the most precise 
         astrometry catalogues at present.}

   \item We predicted in \citet{r:gaia3} that the improvement in accuracy 
         of VLBI and/or Gaia will not reconcile the VLBI and Gaia positions, 
         but will make these differences more significant. This prediction
         has come true. The fraction of outliers grew from 6 to 9\%, and the 
         distribution of the position offset directions as a function of 
         $\psi$ angle became sharper.

   \item We demonstrated that the main reason for the statistically significant 
         VLBI/Gaia position offset is the presence of optical structure. 
         Among the matched sources with the normalized arc lengths 
         exceeding 4 that have measured jet directions, 52\%--62\%, i.e., 
         {\it the majority}, have the position offsets parallel to the 
         jet direction. Therefore, we conclude that the optical jet is the cause.
         Although this fraction may be somewhat lower for the entire population 
         of matched AGNs, we got its firm lower limit: 27\%. Other reasons
         mentioned by \mbox{\citet{r:gaia_dr2_crf}} can explain only a small
         fraction of outliers. 

           The presence of emission from a host galaxy within the Gaia 
         point spread function may shift the centroid with respect 
         to the nucleus if the galaxy central region structure is asymmetric 
         or the AGN is dislodged with respect to the galaxy center of mass, 
         \Note{we assume such a shift is independent on jet direction 
         angle in the absence of evidence of such a dependency}. 
         Table~\ref{t:hist_fit} provides the upper limit of the fraction 
         of outliers which position offsets do not depend on $\psi$: 33\%. 
         It does not seem likely that all of these offsets are caused by 
         the contribution of host galaxies, because the fraction of AGNs 
         with discernible host galaxies is much less. 
         
   \item We found that scaling the Gaia position uncertainties by 
         $\sqrt{\chi^2/\mbox{ndf}}$ eliminated the dependence of the fraction 
         of outliers on $\chi^2/\mbox{ndf}$. Examining the 
         subset of matches with dominating VLBI or Gaia errors allowed us to 
         evaluate the scaling factors for the VLBI uncertainties, 1.30, and 
         the Gaia position uncertainties: $1.06 \, \sqrt{\chi^2/\mbox{ndf}}$. 
         Eliminating the observations within 0.5~rad of $\psi=0$ and 
         $\psi=\pi$ and using re-scaled uncertainties, made the distribution 
         of normalized VLBI/Gaia arc-lengths much closer to the Rayleigh 
         distribution: compare Figures~\ref{f:norm_arc_all} and 
         \ref{f:norm_arc_off}.

   \item The contribution of VLBI and/or Gaia systematic errors on estimates 
         of the orientation angles of the Gaia DR2 catalogue with respect 
         to the VLBI catalogue does not exceed 0.02~mas.

   \item We predicted in \citet{r:gaia3} that flares in AGNs would
         cause a jitter in their positions \Note{because an increase of flux
         in one of the components of an extended source will change the 
         centroid position}. The analysis of Gaia proper 
         motions provided us an indirect confirmation of this prediction: 
         the sources with excessive Gaia residuals, i.e., large $\chi^2$/ndf, 
         have proper motion directions predominately parallel to the jet 
         directions. The median magnitude of statistically significant proper 
         motions is larger than 1~mas/yr over a 1.16~year interval, which is 
         significantly higher than $<0.05$~mas/yr over 5~years anticipated 
         before the Gaia launch \citep{r:perryman14}. Although AGNs proper 
         motions should not be interpreted as a bulk tangential motion, at the 
         same time, these proper motions are not always artifacts of 
         Gaia data analysis. The photo-centers of at least some quasars are 
         not fixed points and the possibility of quasar proper motion should 
         be taken into account in interpreting results of differential 
         astrometry.

   \item We found that VLBI proper motions have a preferable direction 
         along with the jet. Median VLBI proper motions of AGNs are 
         a factor of 50 smaller than Gaia proper motions.

\end{enumerate}

  We do not claim that we have solved the problem of establishing 
the nature of {\it all} outliers. The distribution in 
Figure~\ref{f:norm_arc_off} still deviates from Rayleigh and we still 
did not uncover the nature of the 1/3 statistically significant offsets, 
but we made a quite substantial progress. We anticipate that a study 
of VLBI/Gaia position offsets will become a power tool for probing 
properties of the accretion disk and the relativistic jet in the AGNs, 
in line with the work of \citet{r:gaia5}.

\section*{Acknowledgments}

  We used in our work the Astrogeo VLBI FITS image 
database\footnote{Available at \href{http://astrogeo.org/vlbi_images}
{http://astrogeo.org/vlbi\_images}} that contains radio images 
contributed by A.~Bertarini, L.~Garcia, N.~Corey, Y.~Cui, L.~Gurvits, X.~He,
D.~Homan, S.~Jorstad, S.~Lee, R.~Lico, M.~Lister, E.~Liuzzo, A.~Marscher, 
C.~Marvin, A.~B.~Pushkarev, E.~Ros, T.~Savolainen, K.~Sokolovski, G.~Taylor, 
A.~de Witt, M.~Xu, B.~Zhang, and the authors.
It is our pleasure to thank Eduardo Ros for suggestions that led to
improvements of the manuscript. 

   This project is supported by the Russian Science Foundation grant 16-12-10481.
This work has made use of data from the European Space Agency (ESA) mission 
Gaia
processed by the Gaia Data Processing and Analysis Consortium.
Funding for the DPAC has been provided by national institutions, in particular 
the institutions participating in the Gaia Multilateral Agreement.
We used in our work VLBA data provided by the Long Baseline Observatory that 
is a facility of the National Science Foundation operated under cooperative 
agreement by Associated Universities, Inc. This research has made use of the 
SIMBAD database, operated at CDS, Strasbourg, France.

\bibliographystyle{mnras}
\bibliography{gaia4}

\bsp    
\label{lastpage}

\end{document}